\documentclass[aps,pre,twocolumn,longbibliography]{revtex4-2}

\usepackage{epsfig}
\usepackage{amsmath,amssymb,bm}
\usepackage{graphicx}
\usepackage{verbatim}
\usepackage{mathtools}
\def\ER{Erd\H{o}s-R\'enyi }

\usepackage{xcolor}

\graphicspath{ {./Images/}}

\begin{document}


\title{Effect of initial infection size on network SIR model}

\author{G.~Machado}
\affiliation{Department of Physics, University of Aveiro $\&$ I3N, Campus Universit\'ario de Santiago, 3810-193 Aveiro, Portugal}

\author{G.~J.~Baxter}
\affiliation{Department of Physics, University of Aveiro $\&$ I3N, Campus Universit\'ario de Santiago, 3810-193 Aveiro, Portugal}





\date{\today}

\begin{abstract}
We consider the effect of a nonvanishing fraction of initially infected nodes (seeds) on the SIR epidemic model on random networks. This is relevant when, for example, the number of arriving infected individuals is large, but also to the modeling of a large number of infected individuals, but also to more general situations such as the spread of ideas in the presence of publicity campaigns.
This model is frequently studied by mapping to a bond percolation problem, in which edges in the network are occupied with the probability, $p$, of eventual infection along an edge connecting an infected individual to a susceptible neighbor. 
This approach allows one to calculate the total final size of the infection and epidemic threshold in the limit of a vanishingly small seed fraction. 
We show, however, that when the initial infection occupies a nonvanishing fraction $f$ of the network, 
this method yields ambiguous results, as the correspondence between edge occupation and contagion transmission no longer holds.
We propose instead to measure the giant component of recovered individuals within the original contact network. This has an unambiguous interpretation and correctly captures the dependence of the epidemic size on $f$. We give exact equations for the size of the epidemic and the epidemic threshold in the infinite size limit. We observe a second order phase transition as in the original formulation, however with an epidemic threshold which decreases with increasing $f$.
When the seed fraction $f$ tends to zero we recover the standard results.
\end{abstract}


\maketitle



 \section{Introduction \label{intro}}

Compartmental epidemic models 
provide a powerful mathematical tool for predicting and understanding the spread of contagions. 
In these models, the population is divided into several compartments, representing different states of individuals with respect to the disease. 
Perhaps the most well known and among the most simple compartmental model is the SIR model \cite{anderson1992infectious} which describes the spreading of a contagions, such as a diseases, from which individuals recover with immunity. 
Most individuals begin in the susceptible state, and may become Infected upon contact with an infected individual. Infected individuals may in turn recover at a given rate. Hence the individuals progress through these states in the following order:
\begin{equation}\label{SIR}
S \xrightarrow{\beta} I \xrightarrow{\alpha} R\,.
\end{equation}
One of the most important generalizations of such models is to consider the spread to take place on a heterogeneous contact network \cite{keeling2005networks}. 
Infected nodes pass the infection to susceptible neighbors with some rate ${\beta}$, and recover at a rate ${\alpha}$. The infection may spread to only a small number of individuals, or spread across a significant portion of the network. The process ends when no more infected nodes remain, so that all sites are either recovered or still susceptible.

This remains a dynamic and active field of research, with researchers considering numerous aspects of network disease spreading models \cite{pastor2015epidemic}, including, for example, the effects of network structure  \cite{pastor2001epidemic}, degeree correlations  \cite{boguna2002epidemic}, and heterogeneous transmission rates \cite{baxter2021degree}.
Such models, and more complex variations of them, form the basis of current efforts to model and predict the ongoing SARS-CoV-2 epidemic \cite{pais2020predicting, yamana2020projection}.
Usually, theoretical models use the initial condition of a minute fraction of infected individuals,  \cite{pastor2001epidemic, newman2002spread, keeling2005networks, boguna2002epidemic, pastor2015epidemic}, or a single infection \cite{kenah2007network, miller2007epidemic, rogers2015assessing, baxter2021degree}, with all others initially being susceptible.
However some works have recently considered the effect of non-trivial initial conditions
\cite{radicchi2020epidemic}
\cite{schlickeiser2021analytical} in well-mixed populations. These considerations are particularly relevant to modeling of the present 
epidemic, in which simulations may need to startt from an already non-trivial prevalence of the contagion \cite{radicchi2020epidemic}.
 For example, 
 a country or region that is free from a contagion may still face the arrival of numerous travellers or immigrants from regions in which the disease is widespread.
The initial fraction $f$ of infected individuals may have a significant effect on the progression of the epidemic, \cite{radicchi2020epidemic, schlickeiser2021analytical}. 
The effect of initial seed fraction has also been considered in the bootstrap percolation and $k$-core processes \cite{baxter2010bootstrap, baxter2011heterogeneous}.
 The application of such models is not limited solely to diseases, but very similar models can trace the spread of ideas \cite{goffman1964generalization} and other applications, from communication to finance \cite{rodrigues2016application}.    In these cases, too, the initial seeding of the contagion may be (intentionally) widespread, representing media coverage, political campaigns or advertising.

Our aim in this paper is to explore the effect of the fraction, $f$, of randomly chosen initial infections on the SIR epidemic process occurring on a network.
We show that the usual approach of mapping the epidemic size to the network giant component in a bond percolation process  \cite{grassberger1983critical} leads to ambiguous and inaccurate results, except in the limiting case of vanishing initial infected fraction ($f\to0$). 
We show that calculating the giant component of infected sites within the original contact network (retaining all its edges) overcomes this ambiguity, correctly reflecting the dependence of the size of the outbreak on the initially infected fraction, $f$. 
We observe the same continuous phase transition at a well defined epidemic threshold as in the original SIR model. 
However, in our formulation this threshold now depends on the seed fraction $f$.  This allows, for example, the identification of the size of the initial infection required to provoke a large outbreak. Small changes in $f$ can produce significant changes in this threshold. 

In Section \ref{model} we specify the model which we use, and give the general equations for calculating node infection probabilities. In Section \ref{GGCC} we introduce our proposed measure of the outbreak size, and show how it can be calculated based on the percolation mapping. The epidemic threshold is examined in Section \ref{threshold}, and in Section \ref{results} we summarise the behavior with respect to the parameters $f$ and  $p$, the probability of eventual infection along an edge connecting an infected individual to a susceptible neighbor, using the example of \ER networks. Conclusions are presented in Section \ref{conclusions}.


\section{The Model}\label{model}

We now describe the SIR model in detail.
We consider a system of $N$ individuals, each represented by a different vertex of a 
contact network.
Each individual, can be in one of three states at a given time: Susceptible, Infected or Recovered. A Susceptible node 
can only evolve to the Infected state, and an infected node 
 can only evolve to the Recovered state. Once a node is Recovered 
 it can no longer leave this state. Thus the system evolves according to Eq. (\ref{SIR}).

At the start of the process (\(t = 0\)), a fraction \(f\) of individuals, chosen uniformly at random, are set to the infected state. We will refer to these nodes as the seeds. All remaining individuals are initially susceptible.
Considering finite values for $f$ has important implications for the size of the epidemic and how it should be measured.

Each infected node transmits the infection to each of its susceptible neighbors (who then becomes infected), independently, with average rate $\beta$ per unit time. 
Also, each node in the Infected state moves to the Recovered state at rate \(\alpha\) per unit time. 
Thus, the network in the infinite time limit will be constituted solely of susceptible and recovered nodes.

It is useful at this point to define the terms \textit{infection} and \textit{contact} carefully. We will define \textit{contact} as the transmission of the contagion, with rate $\beta$, from an infected node to a neighbor, irrespective of the state of that neighbor. If contact occurs and the neighbor is susceptible in the moment of contact, the neighbour becomes infected, for which we will use the term infection.

We assume a large, sparse, uncorrelated random contact network. Such a network is defined by its degree distribution $P(q)$. The relative fraction of cycles in the network vanishes as the system size tends to infinity. That is, the network is locally tree-like in the large size limit. This allows us to write self consistency equations for all the relevant quantities.

The usual approach is to map this problem to a bond percolation problem \cite{grassberger1983critical}.
Let us define $p$ to be the probability that an infected individual ever transmits the infection to a given susceptible neighbor before recovering.
A node which has become infected at a certain time $t'$ passes the infection to a susceptible neighbor in a very small interval $(t,t+\delta t)$ with probability $\beta \delta t$ if it has not previously recovered. The infected node recovers with probability $ \alpha \delta t$ in each interval of length $\delta t$. Summing over all times $t>t'$ gives
\begin{equation}
p = \frac{\beta}{\alpha + \beta}.
\end{equation}

Now, starting from seed nodes, the infection traverses each edge with probability $p$. We can consider the edges for which this happens as being occupied with probability $p$.
The nodes which can be reached by following occupied edges are the ones that will eventually become infected. 
Thus all the nodes in any connected cluster in the bond percolation problem which contains at least one seed node will be infected, and hence be recovered at the end of the process.
Thus the fraction of nodes that are infected (and end up recovered), is equal to the fraction of nodes in connected clusters in the bond percolation problem that contain at least one seed. 

Using the locally tree-like property of the network, one
 can then calculate the probability that a randomly selected node is recovered at the end of the process as by writing a self consistent equation
\begin{align}
S_{Z} &= f  + (1-f) \sum_{q=1}^{\infty} P(q) \sum_{l=1}^{q}\binom{q}{l}  Z^q (1-Z)^{q-l} \nonumber\\
&= f  + (1-f) \sum_{q=1}^{\infty} P(q) [1 - (1-Z)^q]\,. \label{Eq.ZGeneral}%
\end{align}
The first term represents the probability that the node is a seed, which is infected from the start.
 The second term gives the probability that the node is a non-seed node
 which has at least 
 one edge leading to an occupied subtree containing at least one seed node, which occurs with probability $Z$.
We can  calculate $Z$ recursively by solving
\begin{align}
Z &= pf + p(1 - f) \sum_{q=2}^{\infty} \frac{qP(q)}{\langle q \rangle} \sum_{l=1}^{q-1} \binom{q - 1}{l} Z^l (1 - Z)^{q - 1 - l} \nonumber\\
&= p \left[ 1 - (1 - f) \sum_{q=1}^{\infty} \frac{qP(q)}{\langle q \rangle} (1 - Z)^{q - 1} \right].\label{Z_general}
\end{align}
The first term represents the probability of following an occupied edge and encountering a seed.  
The second term regards the probability of following an occupied edge and meeting a non-seed node, of any degree, 
which has at least one other edge (not counting the edge along which we arrived at the node) which satisfies the same condition.
These equations depend on the  assumption that the transmission of the disease in each edge is independent of the others, which is valid under the tree-like approximation.


\section{Measuring the Size of the Epidemic}\label{GGCC}

\begin{figure}[h] 
\centering
\includegraphics[width=\columnwidth]{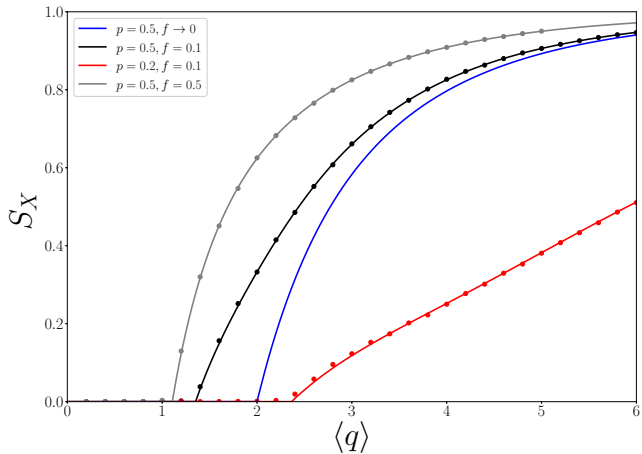}
\caption{Size $S_X$ of the epidemic, measured as the size of the giant component of infected nodes within the original contact network, as a function of mean degree $\langle q\rangle$ for several combinations of seed fraction $f$ and transmission probability $p$. The giant component appears at an epidemic threshold which decreases with increasing $f$, for the same value of $p$. For smaller $p$ the threshold is delayed and the epidemic is smaller. The limit $f\to 0$ corresponds to the usual treatment of the SIR model on a network. Results are shown for \ER networks with $N=10^5$ nodes, averaged over $10$ realisations.
} \label{Fig_examples}
\end{figure}

One may then ask, when does a large outbreak, an epidemic, appear? Typically in network SIR models, one looks for the appearance of a giant component in the percolation  mapping  \cite{newman2002spread}.
This approach gives an unambiguous result when the initially infected nodes form a vanishing fraction of the network ($f\to0$). In this limit, the existence of an occupied edge in the percolation problem represents infection, in clusters where there is a seed.
Then the giant component in the bond percolation problem, if it exists, differs from the entire set of infected nodes only by a small number of infected nodes in finite clusters. One finds a sharp epidemic threshold in the control parameter (network mean degree, say) above which an epidemic appears.

\begin{figure}[h] 
\centering
\includegraphics[width=\columnwidth]{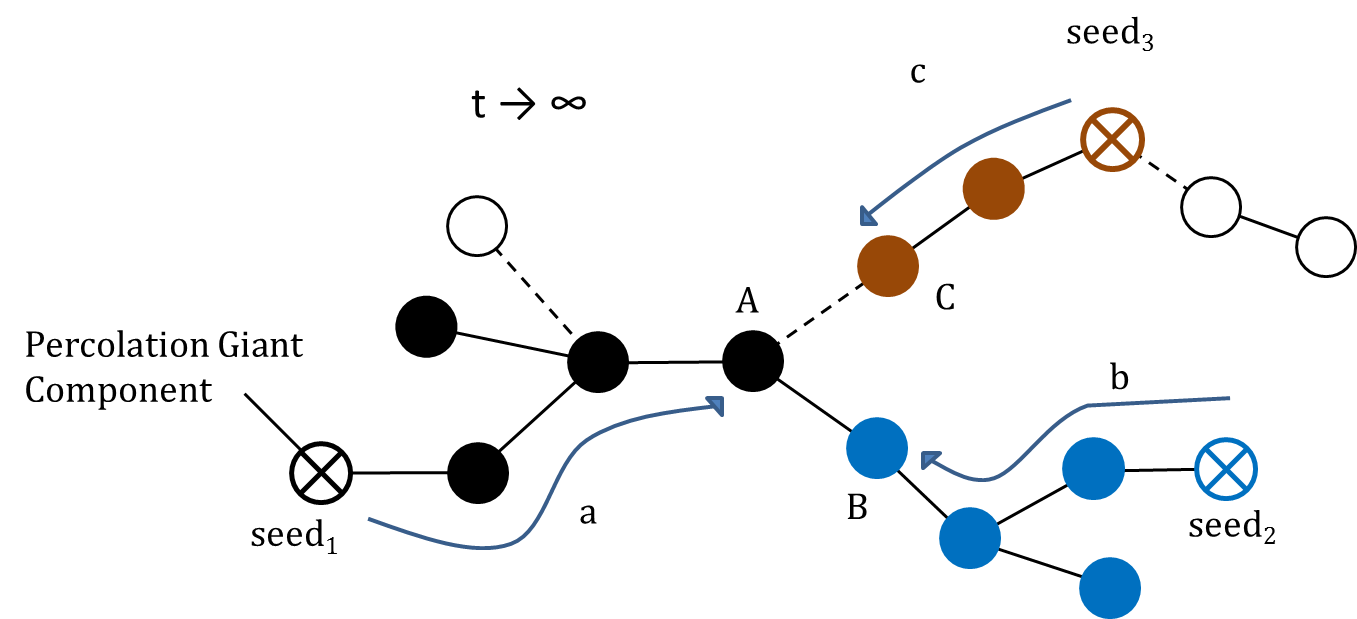}
\caption{Example of how nodes infected from different infection paths can be included/excluded from the percolation giant component due to ambiguous edges. Seed nodes (initial infection) are marked with a cross, and infected nodes are shaded. Arrows indicate the chains of infections. Occupied edges in the percolation mapping are shown as solid lines, unoccupied edges are dashed. 
Node $A$ received the infection via a chain of infection starting from Seed 1, $B$ from that starting at Seed 2, and node $C$ from that starting at Seed 3. Seed 1 has a connection leading to the giant component of the percolation network.
Neither the occupied edge between $A$ and $B$, nor the unoccupied edge between $A$ and $C$  ever passed the infection, in either direction, 
yet The cluster of nodes with infection originating in Seed 2, are connected to the percolation giant component by a path of occupied edges, while the cluster 
with infection originating in Seed 3, is not. 
}  \label{G-GCC.Example}
\end{figure}

When $f>0$, this no longer holds true, and several difficulties arise with the percolation mapping treatment. The non vanishing density of seeds means that occupied edges do not necessarily correspond to the path of infection as there may be multiple  such paths leading to the same node. Furthermore, 
clusters of infections originating from individual seeds may be arbitrarily included or excluded from the percolation giant component, depending on the presence or absence of a connecting occupied edge.
This means that the size of the percolation giant component no longer matches the epidemic size. 
furthermore, as a finite fraction of the network is always infected, there is no longer a sharp transition in the total fraction of infected nodes.
Finally, the percolation giant component size does not change with $f$.
Some of these ambiguities are illustrated in Fig. \ref{G-GCC.Example}.

Our main consideration when addressing this problem is coherence. 
One should either maintain only edges across which the infection was transmitted (severing both links $B$--$A$ and $A$--$C$ in the example shown in Fig. \ref{G-GCC.Example}), or include all adjacent infected components as part of the same cluster (keeping both of the edges in the example).
The first choice would result in a separate connected component for each seed node, creating numerous finite clusters, and would be insufficient to describe the general behavior of the system, \cite{TeseGui} and would make comparison of results for different values of $f$ difficult.
Instead, we choose to measure the size of the giant component of recovered nodes (that have been infected at some time) connected by the original edges of the full contact network, that is, ignoring the status of edges as having been used in infecting or not. Note that this much more closely resembles the situation one might find oneself in when dealing with a real epidemic. In the remainder of this Section we will simply refer to this as the ``giant component".
As we will see, we can still calculate the size of this giant component using the edge occupation probability $p$, through careful formulation of the self consistency equations. We find that the second-order phase transition at the emergence of the giant component is retained, but that the threshold and size of the giant component now depends on the seed fraction $f$. See Fig. \ref{Fig_examples}.

\begin{figure}[h] 
\centering
\includegraphics[width=\columnwidth]{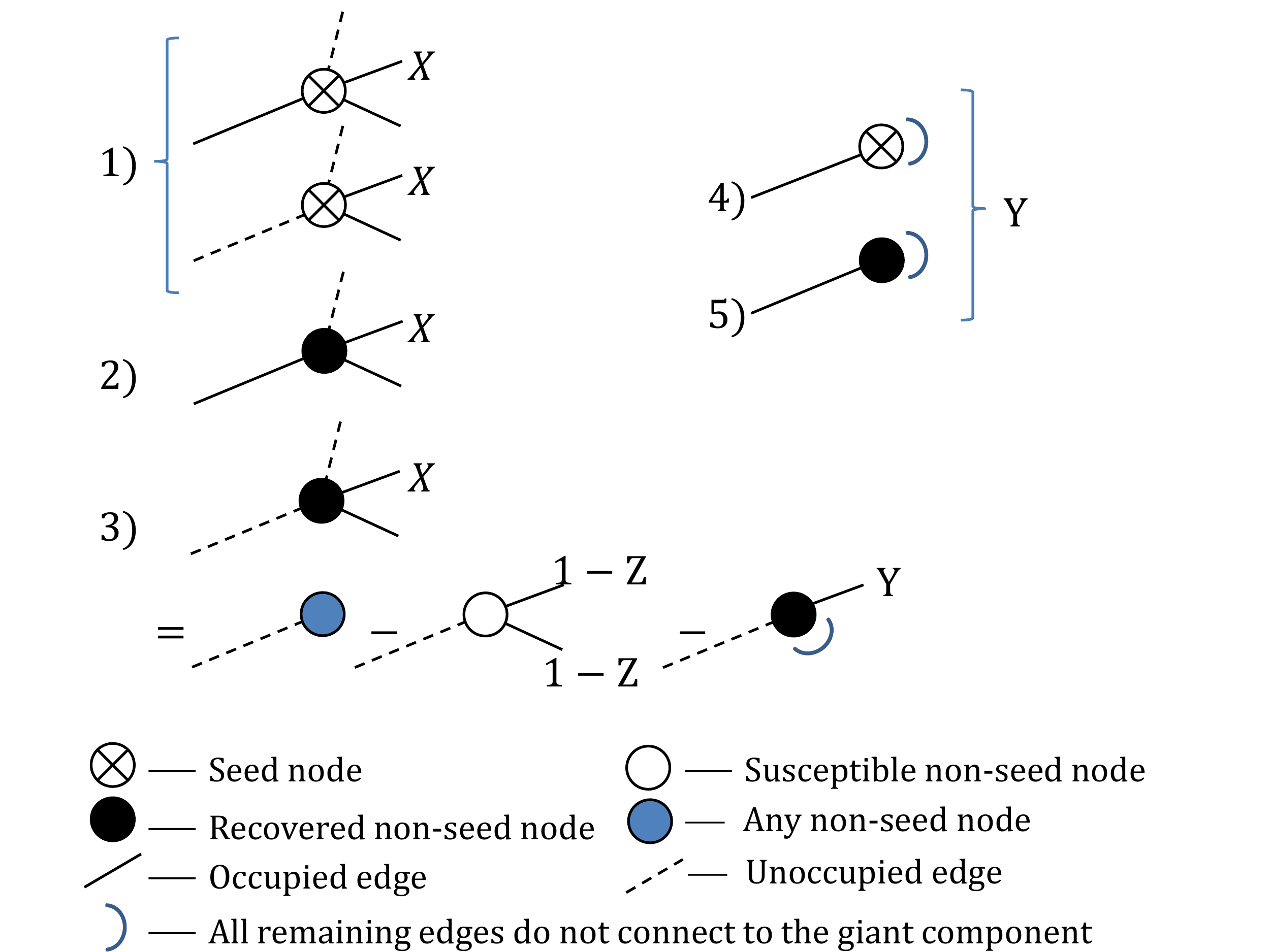}
\caption{All the possible cases contributing to the probability $X$. This occurs when, on following an edge, we encounter a node that was infected at some time, and has at least one further connection satisfying the same condition (indicated by the symbol $X$ in the figure). Numbers $1)$, $2)$ and $3)$ correspond to the terms described in the text. Case $3)$ can be described negatively by also using the probability $Y$, that a node is infected but does not connect to the giant component. The terms contributing to $Y$ are labelled $4)$ and $5)$.
} \label{CasesXg}
\end{figure}

The percolation mapping is still useful for calculating the probability that a given node is ever infected. We therefore look for a giant component of nodes, within the original contact network, for each of which there is a path consisting occupied edges leading to a seed node.

The fraction, \(S_{X}\), of nodes that belong to the giant component 
is equal to the probability that a randomly selected node has at least one connection to the giant component (of recovered nodes within the contact network in the final state), and has at least one connection leading, via a path of occupied edges, to a seed node. This probability thus obeys the following self-consistency equation:
\begin{multline}\label{SXG_main}
S_{X} \! = \!  f \! \sum_{q=1}^{\infty} \! P(q) \! \sum_{l=1}^{q} \! \binom{q}{l} X^l (1 \! - \! X)^{q-l} \! \\
+ \! (1-f) \! \sum_{q=1}^{\infty} \! P(q)  \! \left[1 \! - \! (1 \! - \! Z)^q \! - \! \sum_{l=1}^{q} \! \binom{q}{l} Y^l (1 \! - \! X \! -  \! Y)^{q-l} \right] .
\end{multline}
The first term represents the probability that the node is itself a seed, in which case we just need the condition that it has at least one connection to the giant component, which probability we define as $X$.
The second term, for the case that the node is not a seed,
we construct negatively, by subtracting from one the probability of never being infected [\((1-Z)^q\)]
 and further subtracting the probability that any path leading to a seed node does not lead to the giant component ($Y$), and that none of the other edges lead to the giant component either .

We define \(X\) to be the probability that, starting from a node that is recovered in the final state, and following a random edge, we encounter a neighbor that connects to the giant component via one of its remaining edges (whether occupied or not in the percolation mapping). 
The possible cases in which this can happen, also illustrated in Figure \ref{CasesXg}, are the following:

1) - The edge is either occupied or unoccupied in the percolation mapping, and the node we meet is a seed,  which connects to the giant component by one of its other edges. The probability for this to occur is:
\begin{align}
f\sum_{q=2}^{\infty} \frac{qP(q)}{\langle q \rangle}\sum_{l=1}^{q-1} \binom{q - 1}{l} X^l (1-X)^{q-1-l} ;
\end{align}

2) - The edge is occupied, the node we meet is not a seed, but it connects via one of its remaining edges to the giant component. Notice that the node we meet necessarily ends up recovered, since we are defining \(X\) as coming from a recovered node. Probabilistically this term reads: 
\begin{align}
p(1-f)\sum_{q=2}^{\infty} \frac{qP(q)}{\langle q \rangle} \sum_{l=1}^{q-1} \binom{q - 1}{l} X^l (1-X)^{q-1-l} ;
\end{align}

3) - The edge we are following is unoccupied, the node we meet is not a seed, but it got infected via a different node (than the originating one), and it connects to the giant component. 

This term is  difficult to construct, since it is no longer a subset of the cases of \(Z\). To solve this we construct the possibilities that by following an unoccupied edge, one meets a non-seed node which does {\em not} connect to the giant component, and then subtract them from one. The only cases in which this happens are if either the non-seed node we meet is never infected (all its ongoing edges are cases of \(1 - Z\)), or they correspond to $Z$ but do not connect to the giant component.
We thus introduce \(Y\), the probability that by following an edge from a recovered node, one meets a neighbor that transmitted it the disease, but that does not belong to the giant component. This can happen in two ways, by following an occupied edge that:

4) - leads to a seed, but the seed does not connect to the giant component, with probability 
\begin{align}
pf\sum_{q=1}^{\infty}\frac{qP(q)}{\langle q \rangle} (1-X)^{q-1} ;
\end{align}
or

5) -  that leads to a non-seed, whose further neighbors either did not pass the contagion (unoccupied edge) or do not connect to the giant component. Probabilistically this term reads:
\begin{align}
p(1{-}f) \sum_{q=2}^{\infty} \frac{qP(q)}{\langle q\rangle} \sum_{l=1}^{q-1}\binom{q{-}1}{l} Y^l (1 - X - Y)^{(q-1-l)} .
\end{align}

Notice from Figure \ref{CasesXg}, that  \(Y\) and \(X\) are mutually exclusive.
Thus, combining cases 4) and 5), term 3) will read:
\begin{multline}
(1{-}p)(1{-}f)\sum_{q=2}^{\infty} \frac{qP(q)}{\langle q \rangle} \Big[ 1 - (1 - Z)^{q-1}\\
 - \sum_{l=1}^{q{-}1}\binom{q{-}1}{l}Y^l (1 - X - Y)^{(q-1-l)} \Big].
\end{multline}

Since all the cases that construct \(X\) and \(Y\) are independent of each other, we can simply sum their probabilities,  to obtain:
\begin{align} \label{Eq.XgFull}
X &= 1) + 2) + 3)\\ 
Y &= 4) + 5)\,.
\end{align}
where numbers represent the terms described in the correspondingly numbered paragraphs above.

We can simplify these equations to: 
\begin{multline}
X = \sum_{q=1}^{\infty} \frac{qP(q)}{\langle q \rangle} 
\Big\{
[1 - (1-X)^{q-1}] \\
 + (1{-}p)(1{-}f)  \left[ (1 - X - Y)^{q-1} - (1 - Z)^{q-1}   \right] 
\Big\}  \label{Eq.XgFinalGeneral} 
\end{multline}
and
\begin{multline}
Y = p\sum_{q=1}^{\infty}\frac{qP(q)}{\langle q \rangle} 
\Big[
(1-X)^{q-1} 
- p(1{-}f) (1 - X - Y)^{q-1}
\Big]
\,. \label{Eq.YFinalGeneral}
\end{multline}

Using Eq. (\ref{Eq.ZGeneral}) we can simplify Eq. (\ref{SXG_main}) as
\begin{multline}
S_{X} =  S_{Z} - \sum_{q=0}^{\infty} P(q) \Big[
 (1 - X)^q 
+ (1{-}f) %
 (1- X - Y)^q \Big]. \label{Eq.SxgGeneral}
\end{multline}

One may also write the fraction of recovered 
 nodes that do not belong to the giant component, \(S_{Y}\). Nodes in this situation are:

6) - Seeds that do not connect to the giant component via any of their edges, probability \(f\sum_{q=0}^{\infty} P(q) (1 - X)^q\);

7) - Non-seed nodes with at least one neighbor which transmitted the disease to it, but none of its nieghbors connect to the  giant component, probability \((1-f) \sum_{q=1}^{\infty} P(q) \sum_{l=1}^{q}\binom{q}{l} Y^l (1-X -Y)^{q-l}\). 

Giving
\begin{align}
S_{Y} 
&= \sum_{q=0}^{\infty} P(q) (1 - X)^q - (1-f) \sum_{q=0}^{\infty} P(q) (1-X -Y)^{q} \label{Eq.SyFull} 
\end{align}
so that
\begin{align}
S_{X} = S_{Z} - S_{Y}\,, \label{Eq.GeneralRelation}
\end{align}
with $S_{Z}$ given by Eq. (\ref{Eq.ZGeneral}).


\section{Epidemic threshold}\label{threshold}

Clearly when\(X \ll 1\) ,  \(S_{X} \ll 1\),  and specifically  \(S_{X} = 0\) when $X=0$.
Looking at equations (\ref{Eq.XgFinalGeneral}) and (\ref{Eq.YFinalGeneral}), we can see that the  \(X = 0\) (the probability of connecting to the giant component being zero), is always a solution, (with \(Y\) becoming equal to \(Z\) ).
A second solution for \(X\) eventually appears, marking the emergence of a giant epidemic. To calculate the point at which this happens (the critical point), we assume that the  giant component appears continuously from zero %
and linearize Eq. (\ref{Eq.XgFinalGeneral}) around the critical point:  %
\begin{align} 
X 
& = \frac{\langle q(q{-}1) \rangle}{\langle q \rangle} X - (1{-}p)(1{-}f) G_{2} (1{-}Z ) (X - \delta)
+ \mathcal{O}(X^2)\,, 
\end{align}
where  $\delta \equiv Z-Y$  and
\begin{align}\label{generating}
G_{n}(x) \equiv \sum_{q=0}^{\infty} \frac{P(q)}{\langle q \rangle} \frac{ d^n}{ dx ^n} x^q\,.
\end{align}

Making a similar expansion for $Y$ in orders of $X$ allows us to write an expression for $\delta$ in terms of $X$:
\begin{align}
&\delta = \frac{p \frac{ \langle q(q{-}1) \rangle}{\langle q \rangle}  -  p(1{-}f) G_{2}(1 - Z) }{1 - p(1{-}f) G_{2}(1 - Z)} X\,. \label{delta_G2}
\end{align}
Using Eq. (\ref{delta_G2})  to eliminate $\delta$, eliminating a factor of $X$ and rearranging, we arrive at the condition for the epidemic threshold:
\begin{align}
1 
&= \frac{\langle q(q - 1) \rangle}{\langle q \rangle} \nonumber\\
& - (1{-}p)(1{-}f) G_{2} (1 - Z ) \left[  \frac{ 1 - p \frac{ \langle q(q - 1) \rangle}{\langle q \rangle}}{1 - p(1{-}f) G_{2}(1 - Z)} \right].\label{Eq.CritConditionFinal}
\end{align}
For details of this derivation see Appendix~\ref{Ap.beta}.

Notice that this is not an explicit equation with respect to the model parameters $p$ (itself a function of the rates $\alpha$ and $\beta$), $f$ and the degree distribution $P(q)$ and its moments. Instead one must solve this equation simultaneously with Eq. (\ref{Z_general}). For a specific degree distribution, ane may obtain a closed form condition for the critical point, see the following Section.
Eq. (\ref{Eq.CritConditionFinal}) may be used to obtain the critical point with respect to $p$, for example, holding all other parameters fixed, but also, fixing $p$ (i.e. $\alpha$ and $\beta$)  to find the minimal number of nodes that have to start in the infected state for the emergence of a giant epidemic.

Including further terms in the expansion of Eq. (\ref{Eq.XgFinalGeneral}), we can determine that $X$, and hence the epidemic size $S$, grows linearly above the epidemic threshold:
\begin{equation}
S \approx (p - p_{c})^{\beta}\
\end{equation}
with $\beta = 1$, as for the standard percolation transition. See Appendix \ref{Ap.beta} for the derivation of this result.


\section{Results for \ER Networks\label{results}}

To illustrate the dependence of the epidemic on the various parameters, we now consider the specific case of \ER networks, whose degree distribution is Poisson, $P(q) = \frac{e^{-\lambda} \lambda^q}{q!}$, with mean degree $\langle q\rangle = \lambda$.
In this case, Eqs. (\ref{Eq.ZGeneral}) , (\ref{Eq.SxgGeneral}) and (\ref{Eq.SyFull}) become
\begin{align} 
S_{Z} & = 1 - e^{-\lambda Z},\\
S_{X} &= S_{Z} - e^{-\lambda X}\left[ 1 - (1-f) e^{-\lambda Y} \right] , \label{Eq.Sxg}\\
S_{Y} &= e^{-\lambda X}\left[ 1 - (1-f) e^{-\lambda Y} \right] \label{Eq.Sy},
\end{align}
while Eqs. (\ref{Z_general}), (\ref{Eq.XgFinalGeneral}) and (\ref{Eq.YFinalGeneral}) become
\begin{align} 
Z &= p[1 - (1 - f) e^{- \lambda Z}], \label{Z_ER}\\
X &=  (Z - Y) \left( \frac{1 - p}{p} \right) + p (1 -  e^{-\lambda X}) , \\
Y &=  pe^{-\lambda X} [1 -  (1-f)e^{-\lambda Y}] ,\label{Y_ER} \\
\delta &= \frac{\lambda X Z}{1 - \lambda (p - Z)}.
\end{align}

We illustrate the solution of Eq. (\ref{Eq.Sxg}), which depends on the simultaneous solution of Eqs. (\ref{Z_ER})-(\ref{Y_ER}), in Fig. \ref{Fig_examples}. We see that the size of the epidemic, as well as the threshold at which it emerges, depend on the seed fraction $f$, when other parameters are held equal.

With a Poisson degree distribution as described, we have
\begin{align} 
G_{2}(1 - Z) %
& =\lambda e^{-\lambda Z}
\end{align}
which, substituting into Eq. (\ref{Eq.CritConditionFinal}), gives us the condition for the critical point of an \ER graph: 
\begin{align} 
1 &= \lambda -  (1 - p)(1 - f) \lambda e^{-\lambda Z} \left[1 - \frac{ p \lambda - p (1 - f) \lambda e^{-\lambda Z} }{1 - p (1 - f) \lambda e^{-\lambda Z}} \right]\nonumber\\
&= \lambda + \frac{(1 - p)}{p} \lambda \left( Z - p \right) \left[ \frac{1 - p \lambda}{1 + \lambda (Z - p)} \right].\label{ER_threshcond}
\end{align}
where we have used Eq. (\ref{Z_ER}) to obtain the last line.
Remembering that $Z$ is a function of mean degree $\lambda$, one can obtain the threshold $\lambda_c$ numerically by solving Eq. (\ref{ER_threshcond}) together with Eq. (\ref{Z_ER}).

\begin{figure}[h] 
\centering
\includegraphics[width=\columnwidth]{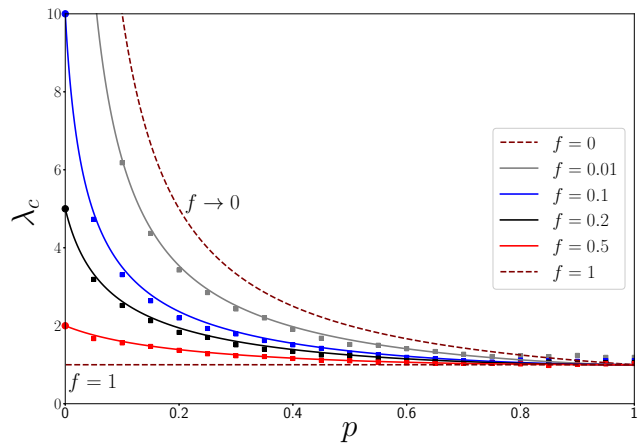} 
\caption{
Epidemic threshold for the mean degree, $\lambda_c $, as a function of $p$ for several values of $f$.
Solid lines represent the theoretical calculation for, from, top to bottom, \(f = 0.01,0.1,0.2,0.5,1\).
The upper and lower dashed lines represent the theoretical calculations for the limits \(f \to 0\) and \(f = 1\), respectively. 
Circular markers indicate the limiting value as \(p \to 0\).
Square markers  represent critical points measured from discrete time simulations on \ER networks of \(N = 10^4\) nodes using 1000 realisations (see Appendix~\ref{Ap.susceptibility}). %
Critical points were identified as the peak value of susceptibility, see Appendix \ref{Ap.susceptibility}.
} \label{Fig.ResultsGGCC}
\end{figure}

Figure \ref{Fig.ResultsGGCC} shows how \(\lambda_c\) changes with \(p\), for different values of \(f\). 
We can see that the critical points for the same \(p\), decrease with increasing $f$, as a greater initial presence of seeds facilitates the formation of the giant component. Similarly, the overall epidemic size is larger for larger $f$ for equal values of all other parameters. See also Fig. \ref{Fig_examples}. If we were to use the standard method of measuring the size of the percolation giant component, we would only obtain the limiting $f=0$ curve. 
It also is observable that, especially for small values, small changes in small values of \(f\) have a large effect on the critical threshold, highlighting the importance of \(f\) on the system's evolution.

In the limit $f=1$, all nodes are seeds, and the epidemic threshold corresponds to the site percolation threshold $\lambda_c=1$, irrespective of the value of $p$. 
In the limit \(p \to 0\), when the seed fraction is not zero, the critical point does not diverge (as it does for $p\to 0$ with $f\to 0$), because the seeds eventually percolate by themselves, at the point \( \langle q \rangle = \frac{1}{f}\), indicated by the circular markers in the figure. 
In the limit \(p = 1\), the giant component of the epidemic corresponds to the giant connected component of the contact network, as all the infected nodes transmit the disease to all their neighbors.

\begin{figure}[h] 
\centering
\includegraphics[width=\columnwidth]{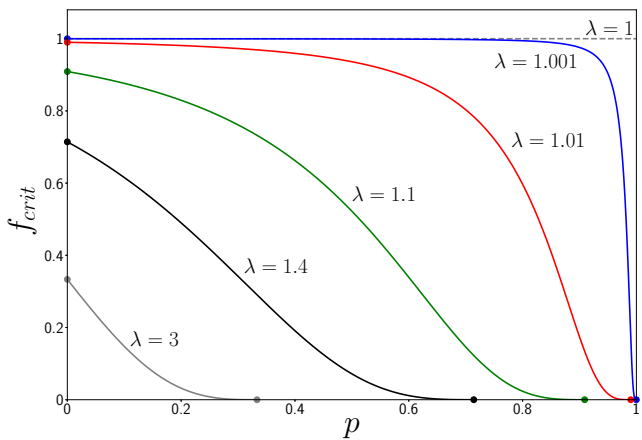} 
\caption{
Epidemic threshold for $f$ as a function of $p$ for several values of $f$.
Solid lines represent the theoretical prediction for the minimal fraction of seeds sufficient for an outbreak to occur, at each value of mean degree \(\lambda\).
The upper dashed line represents the theoretical prediction for the limit \(\lambda \to 1\), the minimal value of \(\lambda\) for the appearance of a Giant Component in the network.
The markers indicate the point at which each line intercepts with each axis.
}  \label{Fig.fwp}
\end{figure}

In Figure \ref{Fig.fwp}, we show the minimal fraction of seeds sufficient for an outbreak to occur (which we define as $f_{crit}$) as a function \(p\), for several values of the mean degree \(\lambda\). 
Naturally, as the probability of infection, \(p\), increases, the minimal fraction of seeds sufficient for an outbreak diminishes.
The superior horizontal dashed line represents the case of \(\lambda = 1\), above which there is no giant epidemic, as there is no giant component of nodes in the contact network in the first place. 
In the limit \(\lambda \to \infty\), \(f_{crit}\) tends to zero, even for \(p \to 0\), as the number of connections between nodes compensates for the lack of seeds and infectiousness.
At the limit \(p = 0\), the threshold corresponds to the percolation threshold of the seeds. 
At the horizontal axis ($f\to 0 $), we find the epidemic threshold for the standard SIR model, for each value of $\lambda$.

\begin{figure}[h]   
\centering
\includegraphics[width=\columnwidth]{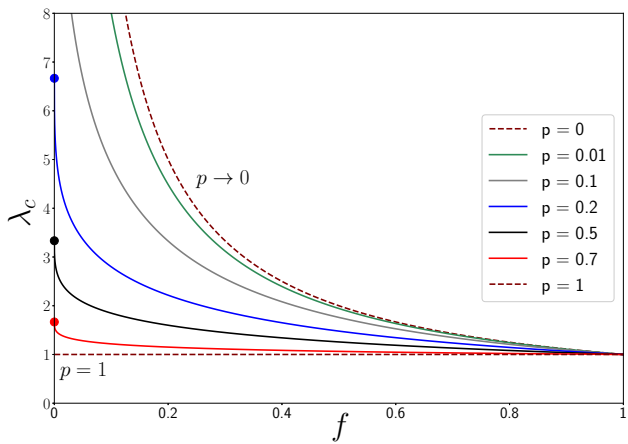} 
\caption{Epidemic threshold $\lambda_c $ as a function of $f$, for several values of $p$.
Solid lines represent the theoretical prediction at each value of $p$.
The upper and lower dashed lines represent the theoretical prediction for the limits \(p \to 0\) and \(p \to 1\), respectively. 
Circular markers indicate the interception between the theoretical lines and the vertical axis, at \(f = 0\).
The lines correspond, from top to bottom, to the values $p = 0,0.01,0.1,0.2,0.5,0.7, 1$.}\label{Fig.Xwf}
\end{figure}

Finally, Figure \ref{Fig.Xwf}, shows how the threshold on the networks' parameter evolves with \(f\), for a fixed \(p\). One may observe that this figure is very similar to figure \ref{Fig.Xwf}. 
These similarities result from the two reasons: firstly, both \(p\) and \(f\) are variables whose increase(decrease) facilitates(hinders) the appearance of an epidemic; 
secondly, there is a symmetry between the limiting curves in both figures.
The lower limit corresponds to the percolation threshold with respect to $\lambda$ of the substrate contact network, when all edges are occupied ($p=1$, Fig. \ref{Fig.Xwf}) or all nodes are occupied as seeds ($f=1$, Fig. \ref{Fig.ResultsGGCC}). Meanwhile the upper limit corresponds to percolation of the seeds for a given value of $\lambda$ ($p\to 0$,  Fig. \ref{Fig.Xwf}) or percolation due to occupied edges ($f\to 0$, Fig. \ref{Fig.ResultsGGCC}).


\medskip

\section{Conclusions \label{conclusions}}

In this paper we have presented a generalization of the well known SIR model of disease spreading on a network, to consider the effect of allowing the fraction of initially infected nodes, $f$, to be nonvanishing. 

The size of the epidemic in the SIR model is usually calculated by mapping the process to a bond percolation problem, and the initial condition is usually taken to be either a single infection, or infection of a vanishingly small fraction of the network ($f\to 0$). In this case, the percolation mapping gives the exact expected outbreak size and epidemic threshold.

We found, however, that allowing the initial fraction of infected nodes, $f$, to be non vanishing leads to a disconnection between the size of the percolation giant component, and the epidemic size. The percolation mapping calculation shows no dependence on $f$. Worse, clusters of recovered nodes may be included or excluded from the giant component arbitrarily, and occupied edges no longer correspond perfectly to disease infection. Thus the calculated giant component does not accurately represent the size of the epidemic or the paths of infection through the network.

To resolve this problem, we measure the giant component on the original network substrate, keeping all ambiguous edges. This mimics the situation one might encounter in a real epidemic: the social contacts are known, but the exact path of the infection may not be, when there are multiple contacts. 
We show how the size of this giant component can be calculated, starting from the percolation mapping. We find that the expected size of the epidemic, and the epidemic threshold both now correctly depend on the initially infected fraction, $f$. 
In particular, we find that a larger initial infection (larger $f$) corresponds to a lower epidemic threshold, or to a larger epidemic when other parameters are held equal.

Near $f=0$, the threshold is highly sensitive to the value of $f$,  showing that an accurate determination of initial conditions is crucial to the correct modeling of the evolution of an epidemic, and correct estimation of other model parameters.

We studied the model on \ER random networks, as an example of sparse random graphs. The results will be qualitatively the same for any uncorrelated random networks with finite first and second moments of the degree distribution. The effect of network topology, such as scale-free network structure, would be an interesting topic for future study.

Our generalization of the SIR model is also relevant to the spread of opinions, ideas, or technology, in which, for example, active campaigning produces an initial seed fraction that is not of negligible size. 
We believe that this work contributes to the general understanding of this fundamental model, and shows that a careful consideration of non-trivial initial conditions should be one element included in more realistic generalisations of the model, and its application to specific situations.


\appendix
\section{Percolation mapping for continuous and discrete time} \label{Ap.p}

Consider the edge connecting an infected node to a susceptible neighbor. The mapping of the contagion process to bond percolation is given by the probability $p$ that the infected node ever infects its susceptible neighbor via this edge \cite{grassberger1983critical}. Here we show how $p$ may be obtained both for a continuous time and discrete time formulation. Once this value is known, all the results given in the main text are valid for either case.

Given transmission rate $\beta$ and recovery rate $\alpha$, the probability that the infection occurs in a very short time interval $\delta t$ is given by 
\begin{equation}
\beta \delta t  (1 - \alpha \delta t).
\end{equation}
The probability that it occurs in a second interval of equal length is, including the probability that it didn't occur in the first interval,
\begin{equation}
\beta \delta t  (1 - a \delta t)(1 - a \delta t)(1 - \beta \delta t)
\end {equation}

Continuing, the probability that the infection occurs in the interval $(t,t+\delta t)$ and not before (where,without loss of generality, we set the time at which the infected node became infected as $t=0$) is
\begin{equation}
P_t = \beta\delta t (1-\alpha\delta t)[(1 - \alpha \delta t) (1 - \beta \delta t)]^{t/\delta t}\,.\label{Pt}
\end{equation}

To obtain the continuous time limit, we take $\delta t \to 0$, obtaining
\begin{equation}
P(t) dt = \beta dt e^{-\alpha t} e^{-\beta t}
\end{equation}
where we have changed to a probability density function $P(t)$.

Integrating $P(t)$ over all times, we obtain the probability that a node ever transmitted the disease before it recovered, \(p\):
\begin{equation}
p = \int_{0}^{\infty}  e^{-\alpha t} e^{- \beta t} \beta dt =  \frac{\beta}{\alpha+\beta}\,.
\label{Eq.pcont}
\end{equation}

For the discrete time formulation, we return to Eq. (\ref {Pt}) and fix the time interval $\delta t \equiv \Delta t$.
The probabilities of infection, $b$, and recovery, $a$, in a single interval are
\begin{align}
a = 1 - e^{- \alpha \Delta t} \label{a_alpha}\\
b = 1 - e^{- \beta \Delta t}\label{b_beta}\,.
\end{align}
We then sum $P_t$ over all time intervals to obtain
\begin{align}
p %
&= \sum_{i = 1}^{\infty} b (1-a)[(1 - a) (1 - b)]^{i-1} = \frac{b(1 - a)}{a + b(1 - a)}\,, 
\label{Eq.pdis}
\end{align}
where $i = t/\Delta t$.

\section{Numerical simulations}\label{Ap.susceptibility}.

Simulation results shown in the figures were generated using discrete time simulations, by choosing single-step infection probability $b$ and recovery probabilities $a$ corresponding to the desired value of $p$. Correspondence to a continuous time formulation can be found by choosing a small value of $\Delta t$ and using Eqs. (\ref{a_alpha}) and (\ref{b_beta}).

In simulations with discrete time, the order of checking for recovery of transmission becomes important.
In our case, we checked for each infected node's recovery first, before they could attempt to transmit the contagion to their neighbors, thus a factor of \(1 - a\) multiplying with \(b\) appears, ensuring that each infected node does not recover in the same timestep it transmits the agent. 
Notice that we recover Eq. (\ref{Eq.pcont}) from Eq. (\ref{Eq.pdis}) when \(\Delta t \to 0\). 

The epidemic threshold is identified numerically by looking for the peak susceptibility, defined as
\begin{align}
\chi = \frac{ \langle S_X^2 \rangle }{ \langle S_X \rangle^2 }
\end{align}
which has no system size dependence, see Ref. \cite{castellano2016numerical} for a discussion of this definition.
In simulations $S_X$ is measured as the size of the largest connected component of recovered nodes at the end of the simulation.

\section{Calculation of the Critical Exponent \(\beta\)} \label{Ap.beta}

Let us define
\begin{align}
G_n(x) \equiv %
\sum_q \frac{P(q)}{\langle q\rangle} \frac{d^n}{dx^n}x^q\,.
\end{align}
Then
\begin{align}
G_1(1) &=  1\\
G_2(1) &= \frac{\langle q(q-1)\rangle}{\langle q\rangle}\\
G_1(1-Z) &=  \sum_q \frac{qP(q)}{\langle q\rangle}(1-Z)^{q-1}\\
G_2(1-Z) &=  \sum_q \frac{qP(q)}{\langle q\rangle}(q-1)(1-Z)^{q-2}
\end{align}
and so on. 
Terms of this sort appear frequently in our equations, so at times it will be useful to write them in terms of $G_n$.
It is also useful to see how these may be expanded in terms of a small perturbation to the parameter $Z$:
\begin{multline}
G_1(1-Z_c-\Delta) \approx G_1(1-Z_c) - G_2(1-Z_c)\Delta Z
\\ + \frac12G_3(1-Z_c)\Delta^2 - ...\label{G1_expansion}
\end{multline}
and
\begin{multline}
G_2(1-Z_c-\Delta) \approx G_2(1-Z_c) - 2G_3(1-Z_c)\Delta  + ...\label{G2_expansion}
\end{multline}
and so on.
We may then rewrite the main equations in terms of $G_n$:
\begin{multline}\label{X_main}
X = 1 - G_1(1-X) - (1-p)(1-f) [G_1(1-Z)
\\ - G_1(1-X-Y)]\,,
\end{multline}
\begin{align}\label{Z_main}
Z %
&= p[G_1(1) -(1-f)G_1(1-Z)]\,,
\end{align}
\begin{align}\label{Y_main}
Y = p[G_1(1-X) - (1-f)G_1(1-X-Y)]\,.
\end{align}
It is clear that for $X=0$, $Y=Z$.

Remembering that $\delta = Z-Y$, 
subtracting Eq. (\ref{Y_main}) from Eq. (\ref{Z_main}) gives
\begin{multline}
\delta = p[1 - G_1(1-X)] \\ - p(1{-}f)[G_1(1-Z) - G_1(1-X-Y)]\,.\label{delta_main}
\end{multline}


Let $p_c$ be the critical value of $p$, at which point $Z$ takes the value $Z_c$.
Writing $p = p_c + \epsilon$ and $Z = Z_c+\Delta$, then near this point, 
Eq. (\ref{Z_main})  may be expanded as
\begin{multline}\label{Z_expansion}
Z_c+\Delta \approx (p_c + \epsilon)[ 1 - (1-f)G_1(1-Z_c)
\\ + G_2(1-Z_c)\Delta - S_3(1-Z_c)\Delta^2]
\end{multline}
which gives
\begin{align}
\Delta \approx \frac{1 - (1-f)G_1(1-Z_c)}{1-p_cG_2(1-Z_c)} \epsilon 
\label{Delta_epsilon}
\end{align}
for $\epsilon \ll 1$, where higher order terms have been neglected.

Similarly, for $Y$, we note that, since $\delta$ is zero at the critical point, $Y_c = Z_c$. So
\begin{align}\label{Y_expansion}
Y = Z -\delta = Z_c + \Delta -\delta 
\end{align}
near the threshold. This means we can use the expansion Eq. (\ref{Z_expansion}), as well as Eqs. (\ref{G1_expansion}) and (\ref{G2_expansion}) for $Y$, by substituting $\Delta-\delta$ for $\Delta$.

We first expand Eq. (\ref{delta_main}) to leading orders of $X$:
\begin{multline}
\delta \approx p[G_2(1)X-G_3(1)X^2] -p(1-f)[G_1(1-Z) 
\\ - G_1(1-Y) + G_2(1-Y)X- \frac12G_3(1-Y)X^2]\,.
\end{multline}
Then we have to expand the generating functions which are functions of $1-Z$ and $1-Y$ in $\Delta$ and $\Delta-\delta$ respectively. Since $Z_c = Y_c$, a few terms cancel, 
and we get, 
neglecting all terms that are second order in small values, 
\begin{align}
\delta \approx pG_2(1)X -p(1-f)G_1(1-Z_c)(X-\delta) .
\end{align}
which can be rearranged to give
\begin{align}\label{delta_smallX}
\delta \approx \frac{pG_2(1) - p(1-f)G_2(1-Z_c)}{1-p(1-f)G_2(1-Z_c)} X\,.
\end{align}
We still haven't considered these expansions in terms of the deviation of $p$ from the critical value $p_c$. We will come to this later.


Now that we have worked out the expansions of all the ``secondary" quantities, we are finally ready to consider the behaviour of $X$ above the critical point.
First we expand the right-hand side of Eq. (\ref{X_main}) in powers of $X$:
\begin{multline}
X \approx G_2(1)X - \frac12G_3(1)X^2
\\- (1{-}p)(1{-}f)\Big[G_1(1{-}Z) \\
- G_1(1{-}Y) + G_2(1{-}Y)X - \frac12 G_3(1{-}Y)X^2
\Big]\,,
\end{multline}
then substitute the expansions for $G_n(1-Z)$ and $G_n(1-Y)$ about $Z_c$, using Eqs. (\ref{G1_expansion}) and (\ref{G2_expansion}):
\begin{multline}
X \approx G_2(1)X - \frac12G_3(1)X^2\\
- (1-p)(1-f) \Big\{
 G_2(1-Z_c)(X-\delta) 
 \\ 
 + 
 \frac12G_3(1-Z_c)\left[
 \Delta^2
 - (\Delta-\delta)^2
- 2X(\Delta-\delta)
- X^2
\right]
\Big\}
\label{X_expansion_final}
\end{multline}
where we have truncated any terms of higher order than $\mathcal{O}(X^2)$ or equivalent.

We first use this expression to obtain the criterion for the critical point. Keeping only terms linear in $X$, substituting the RHS of Eq. (\ref{delta_smallX}) for $\delta$, dividing by $X$ and taking the limit $p \to p_c$ ($X \to 0^+$), we have
\begin{multline}
1 
 = 
 G_2(1)\\
- (1-p_c)(1-f)
 G_2(1-Z_c) \left\{\frac{1 -p_cG_2(1)}{1-p_c(1-f)G_2(1-Z_c)} 
\right\}
\label{critical_criterion}
\end{multline}
which is Eq. (\ref{Eq.CritConditionFinal}).

Now, to find the behavior of $X$ just above this transition point, we write $p = p_c +\epsilon$ in Eq. (\ref{X_expansion_final}). For compactness writing the RHS of Eq. (\ref{delta_smallX}) as $cX$ and the RHS of Eq. (\ref{Delta_epsilon}) as $d\epsilon$. Then
\begin{multline}
X \approx G_2(1)X - \frac12G_3(1)X^2 - (1-p_c-\epsilon)(1-f) \\
\times\Big\{
 G_2(1-Z_c)(1-c)X +
 \\
 \frac12G_3(1-Z_c)\left[
 d^2\epsilon^2
 - (d\epsilon-cX)^2
- 2X(d\epsilon-cX)
- X^2
\right]
\Big\}.
\end{multline}
using Eq. (\ref{critical_criterion}), removing the common factor $X$, and rearranging, we find
\begin{multline}
X
\cong
\frac{2
(1{-}f)(1-c)
\left[ (1{-}p_c)
 G_3(1{-}Z_c) d
+ G_2(1{-}Z_c)\right]
}
{
\left[
G_3(1)
- (1{-}p_c)(1-f) 
G_3(1{-}Z_c)
 (1-c)^2\right]}
 \epsilon
 \label{Xproptoeps}
\end{multline}
where
\begin{align}
1-c &= \left\{\frac{1 -p_cG_2(1)}{1-p_c(1{-}f)G_2(1-Z_c)} \right\} \,,\\
d &= \frac{1 - (1{-}f)G_1(1-Z_c)}{1-p_cG_2(1-Z_c)} \,.
\end{align}
Thus, $X$ grows linearly with the distance above the critical point.

Expanding Eq. (\ref{SXG_main}) in $X \ll 1$ we find
\begin{equation}
S \approx \sum_q P(q) \left\{
qX - (1-f)q(1-Y)^{q-1}(X-\delta)
\right\}\,,
\end{equation}
where we have used that $Z = Y+\delta$. Writing the RHS of Eq. (\ref{delta_smallX}) as $gX$ we have
\begin{align}
S \approx X \sum_q q P(q) \left\{1 - (1-f)q(1-Y)^{q-1}(1-g)\right\}\,.
\end{align}
In other words, $S$ is proportional to $X$ near the critical point, so that, from Eq. (\ref{Xproptoeps}), we can conclude that
\begin{equation}
S \propto \epsilon = (p-p_c)
\end{equation}
near the critical point.

\begin{acknowledgments}
We warmly thank Alexander Goltsev for helpful discussions and suggestions for improving the manuscript.
This work was support by the i3N associated laboratory LA/P/0037/202, within the scope of the projects UID-B/50025/2020 and UID-P/50025/2020, financed by national funds through the FCT/MEC. This work was also supported by National Funds through FCT, I. P. Project No. IF/00726/2015. G. M. was supported by FCT fellowship BD/2020/08794.
\end{acknowledgments}

\bibliography{Bibliography}

\begin{thebibliography}{21}%
\makeatletter
\providecommand \@ifxundefined [1]{%
 \@ifx{#1\undefined}
}%
\providecommand \@ifnum [1]{%
 \ifnum #1\expandafter \@firstoftwo
 \else \expandafter \@secondoftwo
 \fi
}%
\providecommand \@ifx [1]{%
 \ifx #1\expandafter \@firstoftwo
 \else \expandafter \@secondoftwo
 \fi
}%
\providecommand \natexlab [1]{#1}%
\providecommand \enquote  [1]{``#1''}%
\providecommand \bibnamefont  [1]{#1}%
\providecommand \bibfnamefont [1]{#1}%
\providecommand \citenamefont [1]{#1}%
\providecommand \href@noop [0]{\@secondoftwo}%
\providecommand \href [0]{\begingroup \@sanitize@url \@href}%
\providecommand \@href[1]{\@@startlink{#1}\@@href}%
\providecommand \@@href[1]{\endgroup#1\@@endlink}%
\providecommand \@sanitize@url [0]{\catcode `\\12\catcode `\$12\catcode
  `\&12\catcode `\#12\catcode `\^12\catcode `\_12\catcode `\%12\relax}%
\providecommand \@@startlink[1]{}%
\providecommand \@@endlink[0]{}%
\providecommand \url  [0]{\begingroup\@sanitize@url \@url }%
\providecommand \@url [1]{\endgroup\@href {#1}{\urlprefix }}%
\providecommand \urlprefix  [0]{URL }%
\providecommand \Eprint [0]{\href }%
\providecommand \doibase [0]{https://doi.org/}%
\providecommand \selectlanguage [0]{\@gobble}%
\providecommand \bibinfo  [0]{\@secondoftwo}%
\providecommand \bibfield  [0]{\@secondoftwo}%
\providecommand \translation [1]{[#1]}%
\providecommand \BibitemOpen [0]{}%
\providecommand \bibitemStop [0]{}%
\providecommand \bibitemNoStop [0]{.\EOS\space}%
\providecommand \EOS [0]{\spacefactor3000\relax}%
\providecommand \BibitemShut  [1]{\csname bibitem#1\endcsname}%
\let\auto@bib@innerbib\@empty
\bibitem [{\citenamefont {Anderson}\ \emph {et~al.}(1992)\citenamefont
  {Anderson}, \citenamefont {Anderson},\ and\ \citenamefont
  {May}}]{anderson1992infectious}%
  \BibitemOpen
  \bibfield  {author} {\bibinfo {author} {\bibfnamefont {R.~M.}\ \bibnamefont
  {Anderson}}, \bibinfo {author} {\bibfnamefont {B.}~\bibnamefont {Anderson}},\
  and\ \bibinfo {author} {\bibfnamefont {R.~M.}\ \bibnamefont {May}},\
  }\href@noop {} {\emph {\bibinfo {title} {Infectious diseases of humans:
  dynamics and control}}}\ (\bibinfo  {publisher} {Oxford university press},\
  \bibinfo {year} {1992})\BibitemShut {NoStop}%
\bibitem [{\citenamefont {Keeling}\ and\ \citenamefont
  {Eames}(2005)}]{keeling2005networks}%
  \BibitemOpen
  \bibfield  {author} {\bibinfo {author} {\bibfnamefont {M.~J.}\ \bibnamefont
  {Keeling}}\ and\ \bibinfo {author} {\bibfnamefont {K.~T.}\ \bibnamefont
  {Eames}},\ }\bibfield  {title} {\bibinfo {title} {Networks and epidemic
  models},\ }\href@noop {} {\bibfield  {journal} {\bibinfo  {journal} {Journal
  of the Royal Society Interface}\ }\textbf {\bibinfo {volume} {2}},\ \bibinfo
  {pages} {295} (\bibinfo {year} {2005})}\BibitemShut {NoStop}%
\bibitem [{\citenamefont {Pastor-Satorras}\ \emph {et~al.}(2015)\citenamefont
  {Pastor-Satorras}, \citenamefont {Castellano}, \citenamefont {Van~Mieghem},\
  and\ \citenamefont {Vespignani}}]{pastor2015epidemic}%
  \BibitemOpen
  \bibfield  {author} {\bibinfo {author} {\bibfnamefont {R.}~\bibnamefont
  {Pastor-Satorras}}, \bibinfo {author} {\bibfnamefont {C.}~\bibnamefont
  {Castellano}}, \bibinfo {author} {\bibfnamefont {P.}~\bibnamefont
  {Van~Mieghem}},\ and\ \bibinfo {author} {\bibfnamefont {A.}~\bibnamefont
  {Vespignani}},\ }\bibfield  {title} {\bibinfo {title} {Epidemic processes in
  complex networks},\ }\href@noop {} {\bibfield  {journal} {\bibinfo  {journal}
  {Reviews of modern physics}\ }\textbf {\bibinfo {volume} {87}},\ \bibinfo
  {pages} {925} (\bibinfo {year} {2015})}\BibitemShut {NoStop}%
\bibitem [{\citenamefont {Pastor-Satorras}\ and\ \citenamefont
  {Vespignani}(2001)}]{pastor2001epidemic}%
  \BibitemOpen
  \bibfield  {author} {\bibinfo {author} {\bibfnamefont {R.}~\bibnamefont
  {Pastor-Satorras}}\ and\ \bibinfo {author} {\bibfnamefont {A.}~\bibnamefont
  {Vespignani}},\ }\bibfield  {title} {\bibinfo {title} {Epidemic spreading in
  scale-free networks},\ }\href@noop {} {\bibfield  {journal} {\bibinfo
  {journal} {Physical review letters}\ }\textbf {\bibinfo {volume} {86}},\
  \bibinfo {pages} {3200} (\bibinfo {year} {2001})}\BibitemShut {NoStop}%
\bibitem [{\citenamefont {Bogun{\'a}}\ and\ \citenamefont
  {Pastor-Satorras}(2002)}]{boguna2002epidemic}%
  \BibitemOpen
  \bibfield  {author} {\bibinfo {author} {\bibfnamefont {M.}~\bibnamefont
  {Bogun{\'a}}}\ and\ \bibinfo {author} {\bibfnamefont {R.}~\bibnamefont
  {Pastor-Satorras}},\ }\bibfield  {title} {\bibinfo {title} {Epidemic
  spreading in correlated complex networks},\ }\href@noop {} {\bibfield
  {journal} {\bibinfo  {journal} {Physical Review E}\ }\textbf {\bibinfo
  {volume} {66}},\ \bibinfo {pages} {047104} (\bibinfo {year}
  {2002})}\BibitemShut {NoStop}%
\bibitem [{\citenamefont {Baxter}\ and\ \citenamefont
  {Tim{\'a}r}(2021)}]{baxter2021degree}%
  \BibitemOpen
  \bibfield  {author} {\bibinfo {author} {\bibfnamefont {G.}~\bibnamefont
  {Baxter}}\ and\ \bibinfo {author} {\bibfnamefont {G.}~\bibnamefont
  {Tim{\'a}r}},\ }\bibfield  {title} {\bibinfo {title} {Degree dependent
  transmission rates in epidemic processes},\ }\href@noop {} {\bibfield
  {journal} {\bibinfo  {journal} {Journal of Statistical Mechanics: Theory and
  Experiment}\ }\textbf {\bibinfo {volume} {2021}},\ \bibinfo {pages} {103501}
  (\bibinfo {year} {2021})}\BibitemShut {NoStop}%
\bibitem [{\citenamefont {Pais}\ and\ \citenamefont
  {Taveira}(2020)}]{pais2020predicting}%
  \BibitemOpen
  \bibfield  {author} {\bibinfo {author} {\bibfnamefont {R.~J.}\ \bibnamefont
  {Pais}}\ and\ \bibinfo {author} {\bibfnamefont {N.}~\bibnamefont {Taveira}},\
  }\bibfield  {title} {\bibinfo {title} {Predicting the evolution and control
  of the covid-19 pandemic in portugal},\ }\href@noop {} {\bibfield  {journal}
  {\bibinfo  {journal} {F1000Research}\ }\textbf {\bibinfo {volume} {9}}
  (\bibinfo {year} {2020})}\BibitemShut {NoStop}%
\bibitem [{\citenamefont {Yamana}\ \emph {et~al.}(2020)\citenamefont {Yamana},
  \citenamefont {Pei},\ and\ \citenamefont {Shaman}}]{yamana2020projection}%
  \BibitemOpen
  \bibfield  {author} {\bibinfo {author} {\bibfnamefont {T.}~\bibnamefont
  {Yamana}}, \bibinfo {author} {\bibfnamefont {S.}~\bibnamefont {Pei}},\ and\
  \bibinfo {author} {\bibfnamefont {J.}~\bibnamefont {Shaman}},\ }\bibfield
  {title} {\bibinfo {title} {Projection of covid-19 cases and deaths in the us
  as individual states re-open may 4, 2020},\ }\href@noop {} {\bibfield
  {journal} {\bibinfo  {journal} {MedRxiv}\ } (\bibinfo {year}
  {2020})}\BibitemShut {NoStop}%
\bibitem [{\citenamefont {Newman}(2002)}]{newman2002spread}%
  \BibitemOpen
  \bibfield  {author} {\bibinfo {author} {\bibfnamefont {M.~E.}\ \bibnamefont
  {Newman}},\ }\bibfield  {title} {\bibinfo {title} {Spread of epidemic disease
  on networks},\ }\href@noop {} {\bibfield  {journal} {\bibinfo  {journal}
  {Physical review E}\ }\textbf {\bibinfo {volume} {66}},\ \bibinfo {pages}
  {016128} (\bibinfo {year} {2002})}\BibitemShut {NoStop}%
\bibitem [{\citenamefont {Kenah}\ and\ \citenamefont
  {Robins}(2007)}]{kenah2007network}%
  \BibitemOpen
  \bibfield  {author} {\bibinfo {author} {\bibfnamefont {E.}~\bibnamefont
  {Kenah}}\ and\ \bibinfo {author} {\bibfnamefont {J.~M.}\ \bibnamefont
  {Robins}},\ }\bibfield  {title} {\bibinfo {title} {Network-based analysis of
  stochastic sir epidemic models with random and proportionate mixing},\
  }\href@noop {} {\bibfield  {journal} {\bibinfo  {journal} {Journal of
  theoretical biology}\ }\textbf {\bibinfo {volume} {249}},\ \bibinfo {pages}
  {706} (\bibinfo {year} {2007})}\BibitemShut {NoStop}%
\bibitem [{\citenamefont {Miller}(2007)}]{miller2007epidemic}%
  \BibitemOpen
  \bibfield  {author} {\bibinfo {author} {\bibfnamefont {J.~C.}\ \bibnamefont
  {Miller}},\ }\bibfield  {title} {\bibinfo {title} {Epidemic size and
  probability in populations with heterogeneous infectivity and
  susceptibility},\ }\href@noop {} {\bibfield  {journal} {\bibinfo  {journal}
  {Physical Review E}\ }\textbf {\bibinfo {volume} {76}},\ \bibinfo {pages}
  {010101} (\bibinfo {year} {2007})}\BibitemShut {NoStop}%
\bibitem [{\citenamefont {Rogers}(2015)}]{rogers2015assessing}%
  \BibitemOpen
  \bibfield  {author} {\bibinfo {author} {\bibfnamefont {T.}~\bibnamefont
  {Rogers}},\ }\bibfield  {title} {\bibinfo {title} {Assessing node risk and
  vulnerability in epidemics on networks},\ }\href@noop {} {\bibfield
  {journal} {\bibinfo  {journal} {EPL (Europhysics Letters)}\ }\textbf
  {\bibinfo {volume} {109}},\ \bibinfo {pages} {28005} (\bibinfo {year}
  {2015})}\BibitemShut {NoStop}%
\bibitem [{\citenamefont {Radicchi}\ and\ \citenamefont
  {Bianconi}(2020)}]{radicchi2020epidemic}%
  \BibitemOpen
  \bibfield  {author} {\bibinfo {author} {\bibfnamefont {F.}~\bibnamefont
  {Radicchi}}\ and\ \bibinfo {author} {\bibfnamefont {G.}~\bibnamefont
  {Bianconi}},\ }\bibfield  {title} {\bibinfo {title} {Epidemic plateau in
  critical susceptible-infected-removed dynamics with nontrivial initial
  conditions},\ }\href@noop {} {\bibfield  {journal} {\bibinfo  {journal}
  {Physical Review E}\ }\textbf {\bibinfo {volume} {102}},\ \bibinfo {pages}
  {052309} (\bibinfo {year} {2020})}\BibitemShut {NoStop}%
\bibitem [{\citenamefont {Schlickeiser}\ and\ \citenamefont
  {Kr{\"o}ger}(2021)}]{schlickeiser2021analytical}%
  \BibitemOpen
  \bibfield  {author} {\bibinfo {author} {\bibfnamefont {R.}~\bibnamefont
  {Schlickeiser}}\ and\ \bibinfo {author} {\bibfnamefont {M.}~\bibnamefont
  {Kr{\"o}ger}},\ }\bibfield  {title} {\bibinfo {title} {Analytical solution of
  the sir-model for the temporal evolution of epidemics: part b. semi-time
  case},\ }\href@noop {} {\bibfield  {journal} {\bibinfo  {journal} {Journal of
  Physics A: Mathematical and Theoretical}\ }\textbf {\bibinfo {volume} {54}},\
  \bibinfo {pages} {175601} (\bibinfo {year} {2021})}\BibitemShut {NoStop}%
\bibitem [{\citenamefont {Baxter}\ \emph {et~al.}(2010)\citenamefont {Baxter},
  \citenamefont {Dorogovtsev}, \citenamefont {Goltsev},\ and\ \citenamefont
  {Mendes}}]{baxter2010bootstrap}%
  \BibitemOpen
  \bibfield  {author} {\bibinfo {author} {\bibfnamefont {G.~J.}\ \bibnamefont
  {Baxter}}, \bibinfo {author} {\bibfnamefont {S.~N.}\ \bibnamefont
  {Dorogovtsev}}, \bibinfo {author} {\bibfnamefont {A.~V.}\ \bibnamefont
  {Goltsev}},\ and\ \bibinfo {author} {\bibfnamefont {J.~F.}\ \bibnamefont
  {Mendes}},\ }\bibfield  {title} {\bibinfo {title} {Bootstrap percolation on
  complex networks},\ }\href@noop {} {\bibfield  {journal} {\bibinfo  {journal}
  {Physical Review E}\ }\textbf {\bibinfo {volume} {82}},\ \bibinfo {pages}
  {011103} (\bibinfo {year} {2010})}\BibitemShut {NoStop}%
\bibitem [{\citenamefont {Baxter}\ \emph {et~al.}(2011)\citenamefont {Baxter},
  \citenamefont {Dorogovtsev}, \citenamefont {Goltsev},\ and\ \citenamefont
  {Mendes}}]{baxter2011heterogeneous}%
  \BibitemOpen
  \bibfield  {author} {\bibinfo {author} {\bibfnamefont {G.~J.}\ \bibnamefont
  {Baxter}}, \bibinfo {author} {\bibfnamefont {S.~N.}\ \bibnamefont
  {Dorogovtsev}}, \bibinfo {author} {\bibfnamefont {A.~V.}\ \bibnamefont
  {Goltsev}},\ and\ \bibinfo {author} {\bibfnamefont {J.~F.}\ \bibnamefont
  {Mendes}},\ }\bibfield  {title} {\bibinfo {title} {Heterogeneous k-core
  versus bootstrap percolation on complex networks},\ }\href@noop {} {\bibfield
   {journal} {\bibinfo  {journal} {Physical Review E}\ }\textbf {\bibinfo
  {volume} {83}},\ \bibinfo {pages} {051134} (\bibinfo {year}
  {2011})}\BibitemShut {NoStop}%
\bibitem [{\citenamefont {Goffman}\ and\ \citenamefont
  {Newill}(1964)}]{goffman1964generalization}%
  \BibitemOpen
  \bibfield  {author} {\bibinfo {author} {\bibfnamefont {W.}~\bibnamefont
  {Goffman}}\ and\ \bibinfo {author} {\bibfnamefont {V.~A.}\ \bibnamefont
  {Newill}},\ }\bibfield  {title} {\bibinfo {title} {Generalization of epidemic
  theory: An application to the transmission of ideas},\ }\href@noop {}
  {\bibfield  {journal} {\bibinfo  {journal} {Nature}\ }\textbf {\bibinfo
  {volume} {204}},\ \bibinfo {pages} {225} (\bibinfo {year}
  {1964})}\BibitemShut {NoStop}%
\bibitem [{\citenamefont {Rodrigues}(2016)}]{rodrigues2016application}%
  \BibitemOpen
  \bibfield  {author} {\bibinfo {author} {\bibfnamefont {H.~S.}\ \bibnamefont
  {Rodrigues}},\ }\bibfield  {title} {\bibinfo {title} {Application of sir
  epidemiological model: new trends},\ }\href@noop {} {\bibfield  {journal}
  {\bibinfo  {journal} {arXiv preprint arXiv:1611.02565}\ } (\bibinfo {year}
  {2016})}\BibitemShut {NoStop}%
\bibitem [{\citenamefont {Grassberger}(1983)}]{grassberger1983critical}%
  \BibitemOpen
  \bibfield  {author} {\bibinfo {author} {\bibfnamefont {P.}~\bibnamefont
  {Grassberger}},\ }\bibfield  {title} {\bibinfo {title} {On the critical
  behavior of the general epidemic process and dynamical percolation},\
  }\href@noop {} {\bibfield  {journal} {\bibinfo  {journal} {Mathematical
  Biosciences}\ }\textbf {\bibinfo {volume} {63}},\ \bibinfo {pages} {157}
  (\bibinfo {year} {1983})}\BibitemShut {NoStop}%
\bibitem [{\citenamefont {Machado}(2019)}]{TeseGui}%
  \BibitemOpen
  \bibfield  {author} {\bibinfo {author} {\bibfnamefont {G.}~\bibnamefont
  {Machado}},\ }\emph {\bibinfo {title} {{Threshold Generalization of a
  Contagion Model on Complex Networks}}},\ \href@noop {} {Master's thesis},\
  \bibinfo  {school} {Faculdade de Ciências da Universidade do Porto},
  \bibinfo {address} {República Portuguesa} (\bibinfo {year}
  {2019})\BibitemShut {NoStop}%
\bibitem [{\citenamefont {Castellano}\ and\ \citenamefont
  {Pastor-Satorras}(2016)}]{castellano2016numerical}%
  \BibitemOpen
  \bibfield  {author} {\bibinfo {author} {\bibfnamefont {C.}~\bibnamefont
  {Castellano}}\ and\ \bibinfo {author} {\bibfnamefont {R.}~\bibnamefont
  {Pastor-Satorras}},\ }\bibfield  {title} {\bibinfo {title} {On the numerical
  study of percolation and epidemic critical properties in networks},\
  }\href@noop {} {\bibfield  {journal} {\bibinfo  {journal} {The European
  Physical Journal B}\ }\textbf {\bibinfo {volume} {89}},\ \bibinfo {pages} {1}
  (\bibinfo {year} {2016})}\BibitemShut {NoStop}%
\end{thebibliography}%

\end{document}